\documentclass[11pt]{article}

\usepackage[preprint]{acl}

\usepackage{times}
\usepackage{latexsym}
\usepackage[T1]{fontenc}
\usepackage[utf8]{inputenc}
\usepackage{microtype}
\usepackage{inconsolata}
\usepackage{graphicx}

\usepackage{amsmath}
\usepackage{amssymb} 
\usepackage{xcolor}
\usepackage{mathtools}
\usepackage[safe]{tipa}
\usepackage[greek,english]{babel}
\newcommand{\myipa}[1]{{\fontencoding{T3}\selectfont #1}}
\usepackage{enumitem}  
\usepackage{float}
\usepackage{booktabs}
\usepackage{multirow}
\usepackage{xspace}
\usepackage{makecell}
\usepackage{url}
\usepackage{bbding}
\usepackage{threeparttable}
\newcommand{\modelname}{X-Voice\xspace}
\newcommand{\stageone}{X-Voice$_{\text{s1}}$\xspace}
\newcommand{\stagetwo}{X-Voice$_{\text{s2}}$\xspace}

\newcommand{\e}{\mathbf{E}}

\newcommand{\gcheck}{\textcolor{green}{\Checkmark}}
\newcommand{\rcross}{\textcolor{red}{\XSolidBrush}}

\title{X-Voice: Enabling Everyone to Speak 30 Languages via Zero-Shot Cross-Lingual Voice Cloning}

\author{
  Rixi Xu$^{1}$\thanks{These authors contributed equally.}, 
  Qingyu Liu$^{1,3}$\footnotemark[1], 
  Haitao Li$^{2,6}$, Yushen Chen$^{1,2}$, Zhikang Niu$^{1,2}$,\\
  \bf{Yunting Yang$^{4}$, Jian Zhao$^{4}$, Ke Li$^{5}$, Berrak Sisman$^{3}$,}\\
  \bf{Qinyuan Cheng$^{2, 7}$, Xipeng Qiu$^{2, 7}$, Kai Yu$^{1}$, Xie Chen$^{1,2}$\thanks{Corresponding author.}} \\
  $^{1}$MoE Key Lab of Artificial Intelligence, X-LANCE Lab, Shanghai Jiao Tong University,\\
  $^{2}$Shanghai Innovation Institute, $^{3}$Center for Language and Speech Processing, \\Johns Hopkins University,
  $^{4}$Geely, $^{5}$Dataocean AI, $^{6}$Zhejiang University, $^{7}$Fudan University\\
  \texttt{\{sunny\_xrxrx, chenxie95\}@sjtu.edu.cn}
}

\begin{document}
\maketitle
\begin{abstract}
In this paper, we present \textbf{\modelname}, a 0.4B multilingual zero-shot voice cloning model that clones arbitrary voices and enables everyone to speak 30 languages. \modelname is trained on a 420K-hour multilingual corpus using the International Phonetic Alphabet (IPA) as a unified representation. 
To eliminate the reliance on prompt text without complex preprocessing like forced alignment, we design a two-stage training paradigm.
In Stage 1, we establish \stageone through standard conditional flow-matching training and use it to synthesize 10K hours of speaker-consistent segments as audio prompts. In Stage 2, we fine-tune on these audio pairs with prompt text masked to derive \stagetwo, which enables zero-shot voice cloning without requiring transcripts of audio prompts.
Architecturally, we extend F5-TTS by implementing a dual-level injection of language identifiers and decoupling and scheduling of Classifier-Free Guidance to facilitate multilingual speech synthesis.  Subjective and objective evaluation results demonstrate that \modelname outperforms existing flow-matching based multilingual systems like LEMAS-TTS and achieves zero-shot cross-lingual cloning capabilities comparable to billion-scale models such as Qwen3-TTS. To facilitate research transparency and community advancement, we open-source all related resources\footnote{ \url{https://github.com/sunnyxrxrx/X-Voice}}.
\end{abstract}

\section{Introduction}
Zero-shot voice cloning has revolutionized text-to-speech (TTS) synthesis~\citep{tacotron, fastspeech, vits, valle, tortoise, basetts, seedtts}, allowing any target speaker's voice to be cloned from a short audio prompt. Recently, an increasing number of studies have investigated multilingual zero-shot TTS to enable cross-lingual synthesis. 
Pioneering work YourTTS~\citep{yourtts} achieves zero-shot cross-lingual transfer via normalizing flows~\citep{normalizingflow}, but it does not scale beyond a handful of languages. 
With the rise of Large Language Models (LLMs), many subsequent systems have realized multilingual speech generation by leveraging the strong capabilities of LLMs. Most of them adopt one of two mainstream designs: either predicting discrete acoustic tokens in a fully autoregressive (AR) architecture, as in VALL-E~X~\citep{vallex}, Fish-Speech~series~\citep{fish, fishs2}, MOSS-TTS~\citep{moss}, and Qwen3-TTS~\citep{qwen3tts}, or employing coarse-to-fine hybrid frameworks that bridge AR semantic modeling with non-autoregressive (NAR) acoustic refinement, as in Minimax-Speech~\citep{minimaxspeech}, Cosyvoice~series~\citep{cosyvoice, cosyvoice2, cosyvoice3} and IndexTTS~series~\citep{indextts2, indextts25}. While these models effectively scale to dozens of languages and provide high-quality synthesis, their autoregressive paradigm suffers from an inescapable inference bottleneck and error accumulation~\citep{improving}.

\begin{figure*}[ht]
  \centering
  \includegraphics[width=0.85\textwidth]{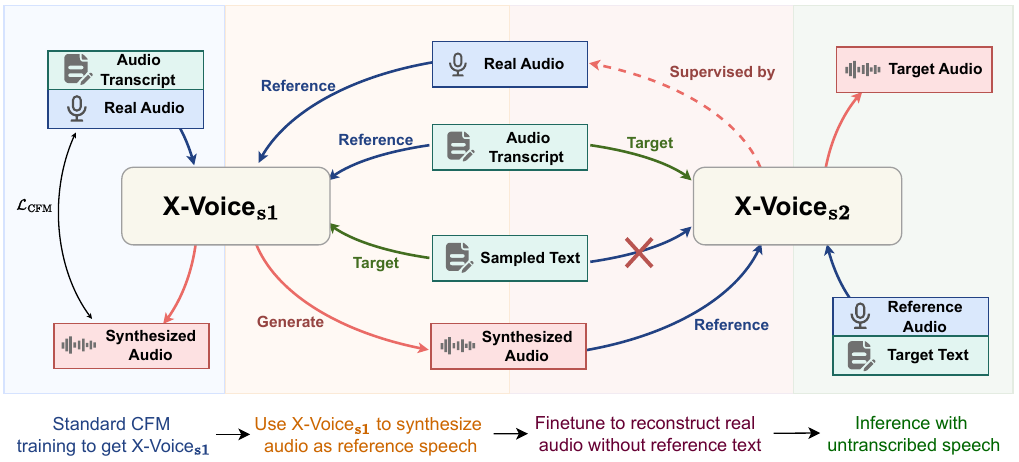}
  \caption{Overview of two-stage training paradigm of \modelname.}
  \label{fig:overview}
\end{figure*}

As an alternative, NAR models mitigate the inherent limitations of AR models and achieve faster inference via parallel decoding. Representative examples include Voicebox~\citep{voicebox}, E2-TTS~\citep{e2tts}, F5-TTS~\citep{f5tts} and NaturalSpeech~series~\cite{naturalspeech,naturalspeech2,naturalspeech3}. Extending these NAR frameworks across languages calls for a unified text representation that enables effective in-context acoustic alignment. While prior work has explored bytes~\citep{byteallyouneed,byte2speech} or graphemes~\citep{grapheme} as input, phonemes remain the most widely adopted representation, as exemplified by Voicebox~\cite{voicebox} and LEMAS-TTS~\cite{lemas}. 

However, the prevailing paradigm in zero-shot voice cloning relies heavily on paired speech and transcript as prompts. While obtaining accurate transcripts is relatively straightforward for high-resource languages, it becomes exceedingly difficult in multilingual settings, particularly for low-resource languages and unwritten dialects. This poses a major bottleneck for cross-lingual speech synthesis, where users may provide spontaneous speech without standardized textual transcripts. To address this inherent limitation, it is essential to eliminate models' reliance on reference transcripts. To this end, recent works have explored several directions: employing speaker encoders to extract identity independently of text~\citep{minimaxspeech, qwen3tts}, utilizing inference-time classifiers~\citep{guidedtts2}, or integrating forced alignment into flow-matching frameworks~\citep{clf5, voxtream2}. However, these approaches either introduce architectural complexity via auxiliary modules or impose heavy pre-processing dependencies that are prone to cascading errors, particularly when scaling to diverse multilingual datasets.

To eliminate the necessity for reference transcripts without introducing structural complexity or pre-processing overhead, and thus enable everyone to speak different languages, we present \textbf{\modelname}, a 0.4B flow-matching model tailored for \textbf{transcript-free cross-lingual voice cloning} in 30 languages. We employ the \textbf{International Phonetic Alphabet (IPA)} as a unified phonetic representation and introduce a \textbf{two-stage training paradigm}, as shown in Figure~\ref{fig:overview}. 
In Stage 1, we build a robust multilingual backbone \stageone based on the F5-TTS~\citep{f5tts} architecture, trained on 420K hours of curated speech.
In Stage 2, we curate a 30K-hour high-fidelity subset and leverage \stageone to synthesize speaker-consistent audios by pairing randomly sampled texts with audios from the subset. These synthetic samples then serve as prompts to reconstruct original real speech. By masking the prompt texts during fine-tuning, we derive \stagetwo, which enables voice cloning independently of reference transcripts.
Moreover, instead of simply merging language identifiers with text tokens, we adopt dual-level language injection (textual level and time level), which effectively alleviates accent leakage in cross-lingual settings. During inference, we introduce a decoupled, scheduled Classifier-Free Guidance (CFG) to improve speech intelligibility and naturalness. 

The main contributions of this paper are summarized as follows:
\begin{itemize}[leftmargin=*, itemsep=-1.5pt, topsep=1pt, partopsep=5pt]
    \item \textbf{Parameter-Efficient Multilingual Foundation:} We present \modelname, a highly efficient 0.4B flow-matching TTS system supporting 30 languages. Through tailored designs like Dual-level Language Injection and Decoupled Classifier-Free Guidance, it achieves high-fidelity zero-shot voice cloning across 30 languages.
    \item \textbf{Open-Source Multilingual Ecosystem:} To facilitate multilingual TTS research, we fully open-source our 420K-hour training corpus and the 30K-hour high-quality subset. Furthermore, we construct and release a rigorously verified benchmark, establishing an evaluation standard for zero-shot multilingual voice cloning.
    \item \textbf{Transcript-Free Supervised Fine-Tuning Paradigm:} We introduce a novel two-stage training pipeline to remove the model's dependence on reference transcripts without relying on auxiliary modules or forced alignment. 
\end{itemize}

\section{Dataset Preparation}
\subsection{Training Corpus}
To establish a robust multilingual foundation, we curate a massive corpus of 420K hours across 30 languages from several open-source datasets. Detailed dataset source and construction process are available in Appendix~\ref{apd:dataset}.
\paragraph{Processing Pipeline} 
To handle noise and inconsistencies in in-the-wild speech data, we implement a rigorous multi-stage processing pipeline to get high-quality training pairs:
\begin{itemize}[leftmargin=*, itemsep=-1.5pt, topsep=1pt, partopsep=5pt]
    \item \textbf{Temporal and Speaking Rate Constraints:} To ensure stable alignment, we prune audio segments shorter than 0.5s or longer than 30s. Then, we calculate the speaking rate (characters per second) for each sample and filter samples using language-specific thresholds.
    \item \textbf{Transcript Language Check:} We employ the langdetect library\footnote{\url{https://github.com/fedelopez77/langdetect}} to verify the linguistic identity of the transcribed text. Samples where the predicted text language conflicts with the source dataset label are discarded. 
    
    \item \textbf{Deduplication Filtering:} Utterances appearing more than 20 times are removed. This process prevents the model from over-fitting to repetitive templates and encourages the model to generalize across a more diverse linguistic distribution.
    \item \textbf{Acoustic Quality Scoring:} We utilize DNSMOS~\citep{dnsmos} to assign an acoustic quality score to every utterance in the corpus. This enables flexible, threshold-based selection for curating high-fidelity data.
\end{itemize}
\paragraph{Dataset Statistics}
\begin{figure}[ht]
    \centering
    \includegraphics[width=\linewidth]{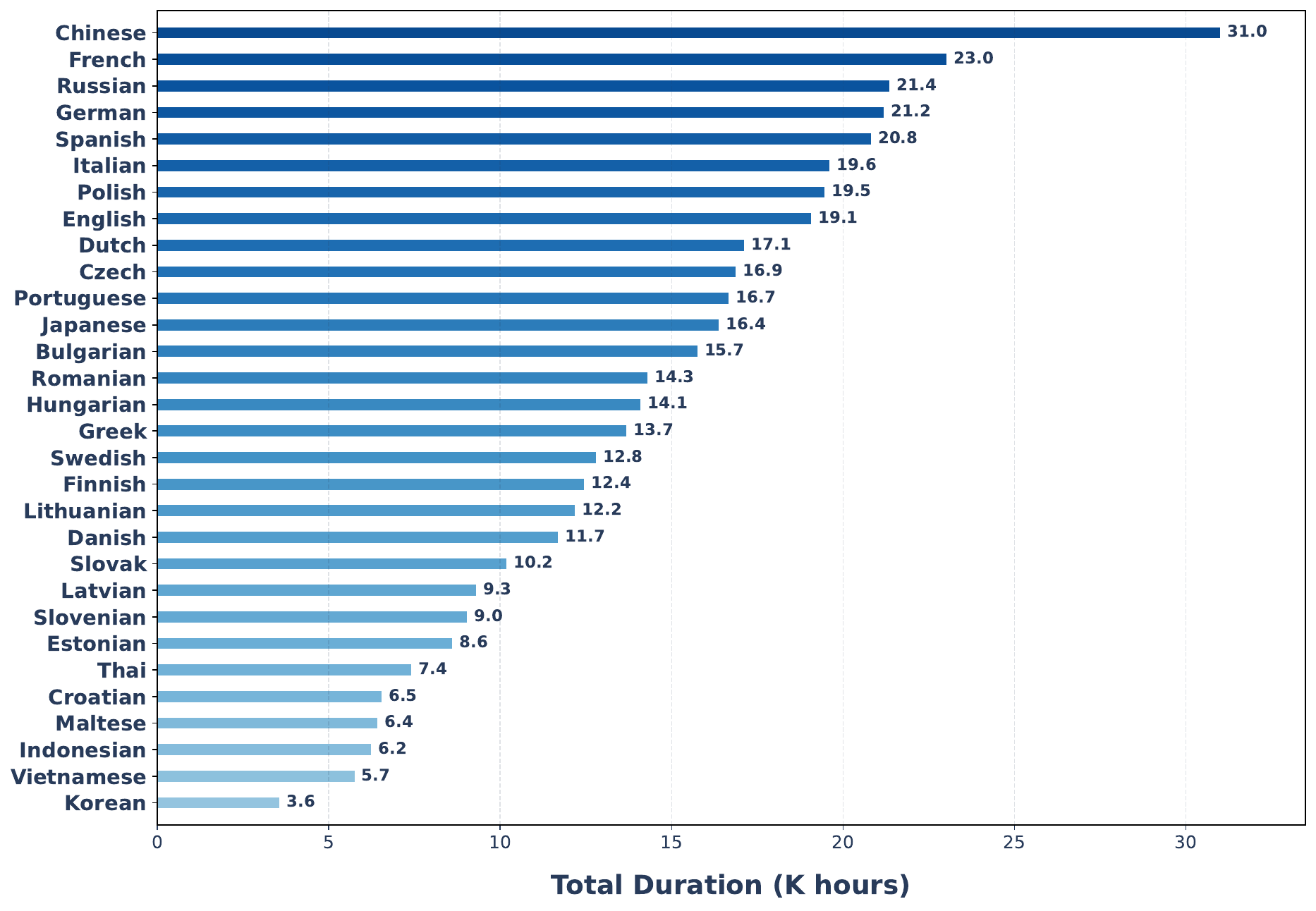}
    \caption{Duration statistics of the X-Voice dataset across 30 languages.}
    \label{fig:total_duration}
\end{figure}
As is shown in Figure~\ref{fig:total_duration}, our finalized corpus integrates 420K hours of speech across 30 languages. The scale and diversity of the dataset enable models to learn both fine-grained phonetic details and higher-level prosodic patterns.

\subsection{Evaluation Benchmark}
Despite the rapid proliferation of zero-shot TTS models, the community still lacks a widely adopted multilingual test set to evaluate these models. To bridge this gap and establish a rigorous benchmark for future research, we construct a high-fidelity test set across 30 languages, which is primarily derived from Common~Voice~\citep{commonvoice}. Given that Common Voice provides limited data for certain languages, the test data for Vietnamese, Korean, and Croatian are collected from Dolly-Audio~\citep{dollyaudio}, Emilia~\citep{emilia}, and ParlaSpeech-HR~\citep{parlaspeech1,parlaspeech2}, respectively. We construct the test set using a multi-stage curation pipeline. As a first step, we filter the test samples by their duration and speaking rate, retaining only well-formed instances. Then, we utilize Silero~VAD\footnote{\url{https://github.com/snakers4/silero-vad}} to detect speech boundaries and trim the leading and trailing silence. This reduces the impact of non-speech segments on evaluation. Finally, to ensure speaker consistency between the reference audio and the ground truth, we employ the ECAPA-TDNN\footnote{\url{https://huggingface.co/speechbrain/spkrec-ecapa-voxceleb/tree/main}}~\citep{ecapa} model and calculate the cosine similarity between the prompt and the ground truth. Only pairs exceeding a similarity threshold of $0.6$ are accepted. For each language, our benchmark contains \textbf{500 human-recorded utterances from over 100 speakers, with ground truth target audios provided.}

We further release a standardized text normalization frontend and evaluation scripts to ensure the fairness and reproducibility of the results.

\section{Methodology}
\subsection{Preliminaries: Flow-matching based DiT}
\modelname is built upon F5-TTS~\citep{f5tts}, a Flow Matching model using Diffusion Transformer (DiT)~\citep{dit} architecture, trained on the text-guided speech-infilling task.
\paragraph{Flow Matching Objective} 
F5-TTS adopts DiT with Optimal Transport Condition Flow Matching (CFM)~\citep{flowmatching}. Let $x_1$ be the Mel-spectrogram and $x_0\in \mathcal{N}(0,I)$. The probability path is defined as $\psi_t(x)=(1-t)x_0+tx_1$. The model $v_t$ is trained to predict the constant vector field $\frac{\text{d}}{\text{d}t}\psi_t(x_0)=x_1-x_0$ by minimizing
\begin{equation}
\small
\mathcal{L}_\text{CFM} = \e_{t, q(x_1), p(x_0)} \left\|  v_t(\psi_t(x_0)) - \frac{\text{d}}{\text{d}t}\psi_t(x_0)\right\| ^2 .
\end{equation}
During inference, the target sample $\psi_1(x_0)$ is generated by using an ODE solver to integrate the predicted vector field $\frac{\text{d}\psi_t(x_0)}{\text{d}t} = v_t(\psi_t(x_0))$ over $t \in [0, 1]$, starting from initial noise $\psi_0(x_0) = x_0$.
\paragraph{Classifier-Free Guidance}
Classifier-Free~Guidance~\citep{cfg} is used to enhance the fidelity and conditioning of the generated speech. During training, the conditioning information $c$ is randomly dropped with a fixed probability. At inference, the guided vector field is computed by linearly extrapolating the conditional and unconditional predictions:
\begin{equation}
\begin{aligned}
v_{t,\text{CFG}} 
= &v_t(\psi_t(x_0); \mathbf{c}) \\
&+ w \left( v_t(\psi_t(x_0); \mathbf{c}) - v_t(\psi_t(x_0)) \right)
\end{aligned}
\end{equation}
where $w$ is the guidance strength. In the context of F5-TTS, the condition $c$ specifically represents the joint prompt consisting of the acoustic condition and the text condition.  A smaller $w$ better preserves the reference timbre but results in reduced pronunciation accuracy. By contrast, a larger $w$ strengthens text alignment but compromises speaker similarity and speech naturalness.

\subsection{Overview}
The design philosophy of \modelname centers on a two-stage training paradigm that progressively builds a multilingual foundation and then extends it to transcript-free synthesis: 
\begin{itemize}[leftmargin=*, itemsep=-1.5pt, topsep=1pt, partopsep=5pt]
\item \textbf{\stageone: Multilingual Foundation.} We establish a robust acoustic manifold by training on a 420K-hour multilingual corpus. This stage focuses on learning universal speech representations and stable cross-lingual cloning. 
\item \textbf{\stagetwo: SFT with Synthetic Prompts.} We adopt a supervised fine-tuning (SFT) strategy following Cross-Lingual~F5-TTS~2~\citep{clf5tts2}, where \stageone is used to generate speech that serves as audio prompts. By fine-tuning on paired synthetic prompts and real target speech, \modelname enables transcript-free voice cloning while preserving synthesis quality.
\end{itemize}
A key distinction of our approach is the complete absence of explicit data alignment throughout both stages, relying on the DiT backbone to implicitly learn latent cross-modal correspondences.

\subsection{Unified Multilingual Representation}
To bridge the cross-linguistic gap, we construct a unified phonetic space.  We represent Mandarin Chinese via Pinyin due to its highly standardized syllabic structure. For other languages, we utilize the \textbf{International Phonetic Alphabet (IPA)}. The IPA tokens are derived via eSpeak-NG\footnote{\url{https://github.com/espeak-ng/espeak-ng}} for most of the languages. However, we find that eSpeak-NG exhibits suboptimal performance in representing certain Asian languages. Therefore, we employ specialized toolkits for these languages—PyThaiNLP~\citep{pythainlp} for Thai, PyOpenJTalk\footnote{\url{https://github.com/r9y9/pyopenjtalk}} for Japanese, and g2pK\footnote{\url{https://github.com/kyubyong/g2pK}} for Korean.

We detail the design of our unified multilingual representation as follows. First, we explicitly include stress markers (\myipa{"}) to resolve lexical and prosodic ambiguities. For instance, in the Greek example in Table~\ref{tab:tokenization}, the position of the stress marker is the sole differentiator between two distinct semantic meanings. Adding this token is essential for maintaining prosodic naturalness in synthesized speech. Second, as shown in the English example in Table~\ref{tab:tokenization}, we decompose phonemes into core articulatory units and suprasegmental modifiers (e.g., length marks (\myipa{:}), aspiration ($^{\text{h}}$), and tonal numbers). This design aims to capture the universal acoustic base while separately modeling the acoustic shifts introduced by modifiers. 

\begin{table}[htbp]
\centering
\caption{Examples of our unified phonetic representation for underlined words. We preserve lexical stress tokens and separate primary articulatory units from suprasegmental modifiers.}
\label{tab:tokenization}
\small
\begin{tabular}{@{}lll@{}}
\toprule
\textbf{Language} & \textbf{Transcript} & \textbf{Token Sequence} \\ \midrule
Greek & \foreignlanguage{greek}{Εσύ \underline{πότε} έρχεσαι.}   & \myipa{\texttt{[p, ", o, t, e]}} \\
& \foreignlanguage{greek}{Δεν πάω \underline{ποτέ}.}  & \myipa{\texttt{[p, o, t, e, "]}} \\ \midrule
English & \textit{\underline{See you}.}  & \myipa{\texttt{[s, ", i, :, j, u, :]}} \\
 & \textit{\underline{Go far}.}  & \myipa{\texttt{[g, oU, f, ", A, :, R]}} \\ \bottomrule
\end{tabular}
\end{table}

As for the integration of articulatory units and modifiers,  \citet{revisiting} verifies that compared with separately embedding articulatory tokens and suprasegmental modifiers, unifying their embedding representations leads to negligible performance differences for intra-lingual and cross-lingual voice cloning in NAR TTS systems. Thus, we directly embed articulatory units and suprasegmental features together in a sequence.

\subsection{\stageone: Robust Foundation Modeling}
\begin{figure}[ht]
    \centering
    \includegraphics[width=0.9\linewidth]{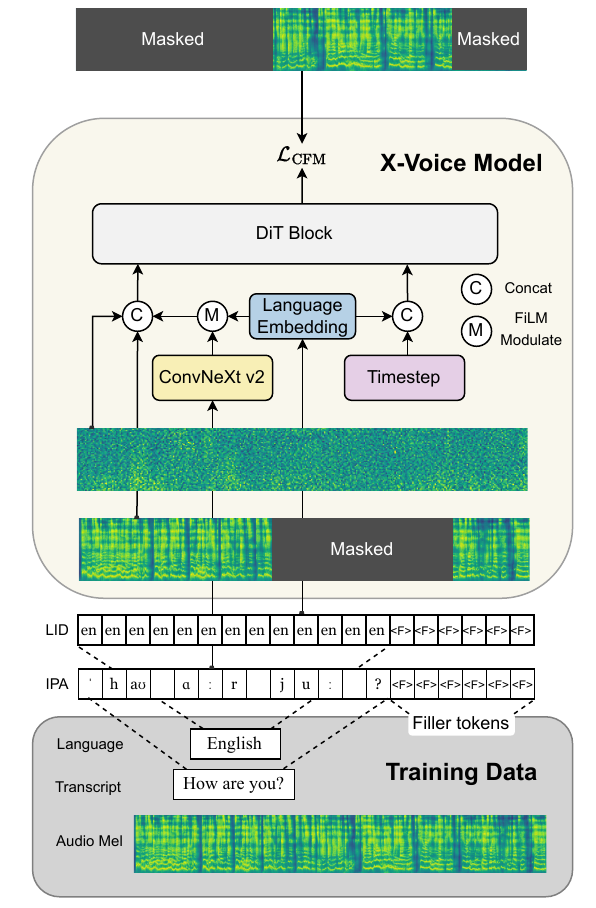}
    \caption{Training Framework of \stageone. We use IPA as unified representation and inject LID in both textual level and time level.}
    \label{fig:stage1}
\end{figure}
\paragraph{Dual-Level Language Injection}
Without explicit language conditioning, large-scale models often struggle to distinguish between different linguistic characteristics, and may lead to accent leakage problem in cross-lingual tasks. Previous work like IndexTTS~2.5~\citep{indextts25} injects Language ID (LID) at  textual level to guide pronunciation. However, our empirical findings suggest that when using textual language conditioning alone, the model still suffers from accent leakage during cross-lingual synthesis. We attribute it to a lack of global constraints to fully decouple the speaker's timbre from their accents. To address this, we propose a \textbf{Dual-Level Language Injection} mechanism that injects LID in both time level and textual level, which provides better pronunciation accuracy and overall prosody.

 At the time level, we propose to inject the LID embedding $\textbf{e}_L$
 by concatenating it with time embedding $\mathbf{e}_t$. This fused vector passes through a Multi-Layer Perceptron (MLP) with SiLU activation:
\begin{equation}
\mathbf{h}_{t} = \text{SiLU}(\mathbf{W} [\mathbf{e}_t \oplus \mathbf{e}_L] + \mathbf{b}).
\end{equation}
This mechanism effectively steers the ODE trajectory to align with the target language's prosodic manifold. 

At the textual level, we identify that text embeddings carry richer information than LID, simply concatenating them may be suboptimal because the sparse LID signal can overshadow phonetic features. We instead apply the Feature-wise Linear Modulation (FiLM)~\citep{film} to text embeddings:
\begin{equation}
\text{FiLM}(\mathbf{e}_T) = \gamma(\mathbf{e}_L) \odot \mathbf{e}_T + \beta(\mathbf{e}_L),
\end{equation}
where $\mathbf{e}_T, \mathbf{e}_L$ are text embedding and LID embedding, $\gamma$ and $\beta$ are learned parameters that scale and shift the phonetic features.
This multiplicative rescaling acts as a gate, forcing the model to adapt shared IPA representations to language-specific acoustic patterns. 

We zero-initialize all LID injection layers to avoid interfering with the pretrained representations at the onset of training.

Our ablation study in Section \ref{sec:lid} demonstrates that omitting this global constraint causes the model to carry over source-language accents during cross-lingual synthesis, whereas the dual-level approach maintains high linguistic purity.

\begin{figure*}[ht]
  \centering
  \includegraphics[width=0.85\textwidth]{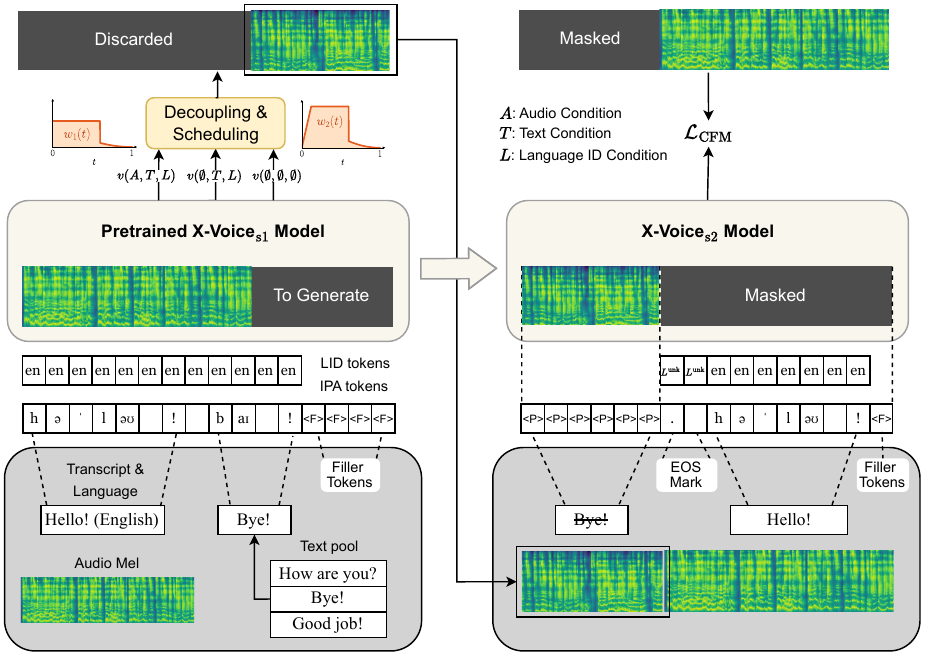}
  \caption{Overview of training paradigm of \stagetwo.
   (Left) \textbf{Synthetic Data Generation:} We utilize the robust \stageone model to synthesize speaker-consistent audio pairs.
    (Right) \textbf{Transcript-Free Fine-tuning:} We fine-tune the model to derive \stagetwo. By replacing reference transcripts with a special placeholder token, we force the model to extract prosodic features directly from the audio prompt without relying on reference transcripts.}
  \label{fig:stage2}
\end{figure*}

\paragraph{Decoupling and Scheduling for CFG}
We have also implemented some optimizations on CFG inference for multilingual scenarios. Existing work mainly focus on the optimization of guidance direction and guidance strength. In terms of guidance direction, \citet{megatts3} and \citet{restyletts} introduce \textbf{Decoupled Classifier-Free Guidance (DCFG)}, which independently steers the \textbf{linguistic} and \textbf{acoustic} components of the vector field for better style control. In terms of guidance strength, \citet{lemas} and \citet{wocfg} introduced a temporal \textbf{decay schedule} to improve naturalness. Inspired by their works, we implement both decoupling and strength decay in our inference process. 

Furthermore, our empirical analysis reveals that the introduction of DCFG intensifies trajectory oscillations during the initial integration steps. This may because abstract text guidance is initially disoriented in high-entropy noise, and strong guidance strength leads to integration shock. By contrast, strong audio prompt guidance is essential for immediately anchoring the speaker's voice outline. Driven by these understandings, we propose \textbf{Asymmetric Warmup (A-Warmup)} for CFG strength, which increases linguistic guidance linearly from zero during the first few inference steps, while maintaining the acoustic guidance at full strength from the onset to lock the acoustic anchor.
Our guided vector field is formulated as
\begin{equation}
\label{eq:dcfg}
\begin{aligned}
     v_{t,\text{DCFG}} &= v_t(\psi_t;A,T,L)\\ 
     &+w_A(t)\cdot\underbrace{(v_t(\psi_t;A,T,L) - v_t(\psi_t;T,L))}_{\text{Acoustic Guidance}}\\
     &+w_L(t)\cdot\underbrace{(v_t(\psi_t;T,L) - v_t(\psi_t))}_{\text{Linguistic Guidance}}, 
\end{aligned}
\end{equation}
where $A, T, L$ denote audio, text, and language conditions, while $w_A, w_L$ denote acoustic and linguistic guidance strengths, respectively. $w_A$ and $w_T$ are functions of time $t$, defined as follows:
\begin{equation}
\label{eq:acsw_weights}
\begin{split}
w_A(t) &=
\begin{cases}
w_A^{\text{start}}, & 0 \le t < t_{\text{decay}}, \\
w_A^{\text{start}} \cdot (1-t)^2, & t_{\text{decay}} \le t < 1.
\end{cases} \\[6pt]
w_L(t) &=
\begin{cases}
w_L^{\text{start}} \cdot \dfrac{t}{t_{\text{warm}}}, & 0 \le t < t_{\text{warm}}, \\
w_L^{\text{start}}, & t_{\text{warm}} \le t < t_{\text{decay}}, \\
w_L^{\text{start}} \cdot (1-t)^2, & t_{\text{decay}} \le t < 1.
\end{cases}
\end{split}
\end{equation}

Scaling $w_A^{\text{start}}$ improves the reference speaker's acoustic similarity, while scaling $w_L^{\text{start}}$ enforces the pronunciation and rhythm of the target language. By tuning them independently, we can maintain comparable speaker similarity while forcefully suppressing the prosodic leakage from the reference audio. The effectiveness of the proposed strategy is demonstrated in the ablation study in Section~\ref{sec:cfg}.

\subsection{\stagetwo: SFT with Synthetic Prompts}
\paragraph{Real–Synthetic Pair Construction}
The goal of synthetic data construction is to generate paired real and synthetic speech for SFT. We filter the original multilingual corpus by ranking samples according to their DNSMOS~\citep{dnsmos} scores, retaining the top 1K hours per language and yielding a high-fidelity subset of 30K hours in total. We then employ \stageone to generate synthetic counterparts for this subset. Specifically, each real sample serves as the audio prompt, while the target text is sampled from a text pool in the corresponding language to drive synthesis. This pipeline produces a total of 10,533 hours of synthetic data for SFT.

\paragraph{SFT without Reference Text}
During SFT, we retain the same text-guided speech-infilling formulation as in \stageone, but remove the reference text associated with the synthetic audio prompt.

The acoustic input is constructed as a mel spectrogram $x_1$ by concatenating the synthetic audio prompt $x_{\text{prompt}}^{\text{syn}}$ with a sequence length $\tau_1$ and the real target speech $x_{\text{target}}$ with a sequence length $\tau_2$:
\begin{equation}
x_1 = [x_{\text{prompt}}^{\text{syn}};\, x_{\text{target}}],
\end{equation}
where $[\cdot;\cdot]$ denotes concatenation along the temporal axis, and $\tau = \tau_1 + \tau_2$ is the total sequence length.

To facilitate the speech infilling task, a binary temporal mask $m$ is introduced and defined as follows:
\begin{equation}
m[:, u] =
\begin{cases}
0, & u \in [0, \tau_1), \\
1, & u \in [\tau_1, \tau).
\end{cases}
\end{equation}
where $u$ indexes mel frames, $(1-m)\odot x_1$ corresponds to the synthetic audio prompt, and $m \odot x_1$ corresponds to the masked target speech.

The text sequence is padded with filler tokens $\langle F\rangle$ to the same length as the mel spectrogram. These filler tokens serve only as padding and carry no semantic information. To reduce the mismatch between the pretraining and SFT text conditions, we prepend $N$ learnable prompt tokens $\langle P\rangle$ with an end-of-sequence (EOS) marker $\texttt{`.\ '}$ before the target text, following Cross-Lingual~F5-TTS~2~\citep{clf5tts2}. The number of prompt text tokens $N$ is determined based on the ratio of the prompt speech duration and the target speech duration. The extended text sequence $z$ is constructed as:
\begin{equation}
\begin{aligned}
z = (&\underbrace{\langle P\rangle,\ldots\ldots, \langle P\rangle}_{N\ \text{times}},\, \texttt{`.'},\,\texttt{`\ '}, \\
     &\, c_1,c_2,\ldots,c_M,\underbrace{\langle F\rangle,\ldots\ldots,\langle F\rangle}_{(\tau-N-M-2)\ \text{times}}),
\end{aligned}
\end{equation}
where $c_i$ denotes the $i$-th token in the target text sequence of length $M$.

To match the dual-level language injection used in \stageone, we apply language conditioning separately to the time embeddings and the text embeddings. For the time embeddings, the global language condition is fixed to the target language embedding $\mathbf{e}_L^{\text{tgt}}$. For the text embeddings, language conditioning is applied selectively over the text sequence at the textual level. The learnable prompt tokens $\langle P\rangle$ and the filler tokens $\langle F\rangle$ are left without any language embedding. The end-of-sequence (EOS) marker \texttt{`.\ '} are associated with a dedicated unknown LID $L^{\text{unk}}$, while the target text tokens $c_{1:M}$ are associated with the target LID $L^{\text{tgt}}$. The textual-level LID sequence $l$ for the text embeddings is defined as:

\begin{equation}
\begin{aligned}
l = (&\underbrace{\varnothing,\ldots\ldots,\varnothing}_{N\ \text{times}},\, {L}^{\text{unk}},\, {L}^{\text{unk}}, \\
     &\, \underbrace{{L}^{\text{tgt}},\ldots\ldots,{L}^{\text{tgt}}}_{M\ \text{times}},\, \underbrace{\varnothing,\ldots\ldots,\varnothing}_{(\tau-N-M-2)\ \text{times}}),
\end{aligned}
\end{equation}
where $\varnothing$ indicates that no LID is injected.

\section{Experiments}
\subsection{Training and Inference Setup}
\paragraph{Training} We follow the same model configurations in F5-TTS~\citep{f5tts} for \modelname and Cross-Lingual~F5-TTS~\citep{clf5} for Speaking Rate Predictor. Detailed model configurations are provided in Appendix~\ref{apd:config}. 

For \stageone training, we initialize the DiT module with the official checkpoint of F5-TTS-v1-Base\footnote{\url{https://huggingface.co/SWivid/F5-TTS/tree/main/F5TTS_v1_Base}} trained on Emilia~\citep{emilia}, and employ the AdamW optimizer~\citep{adamw} for optimization. The model is trained for 600K updates with a batch size of 38,400 audio frames per GPU and bfloat16 mixed precision. The learning rate is linearly warmed up to $7.5\times 10^{-5}$ over the first 20K updates and then decays linearly for the remaining training steps. The DiT module is frozen during the initial 10K update steps.

For \stagetwo training, we initialize the model from the \stageone checkpoint and keep the same training setup. We conduct SFT for a total of 70K update steps.

For Speaking Rate Predictor training, 
we set the batch size to 19,200 audio frames, and the learning rate is warmed up to $2.5\times10^{-4}$ over the first 7.5k updates and then linearly decays.

\paragraph{Inference} We use the Euler ODE solver with 16 NFE steps, linguistic CFG strength 4.0, acoustic CFG strength 2.5, CFG warmup time 0.01 (first 3 NFE steps), decay time 0.6 (last 2 NFE steps), and sway sampling coefficient -1.0. To estimate duration, we use the length ratio of reference text and target text in \stageone, and leverage the speaking rate predictor in \stagetwo.

\subsection{Evaluation Setup}
\paragraph{Baselines} We compare our models with current multilingual TTS systems including Qwen3-TTS~\citep{qwen3tts}, LEMAS-TTS~\citep{lemas}, MOSS-TTS~\citep{moss}, Fish~Audio~S2~\citep{fishs2}, and OmniVoice~\citep{omnivoice}. Their basic configurations are summarized in Table~\ref{tab:model-comparison} and detailed in Appendix~\ref{apd:baseline}.

\begin{table}[ht]
\centering
\small
\caption{Comparison of \modelname with current multilingual zero-shot TTS systems. Arch. denotes the model architecture (AR for autoregressive, NAR for non-autoregressive). Params. denotes the total parameter of the system. Langs. denotes the number of supported languages. Transcript-Free means the model does not require the transcripts of reference audio.}
\label{tab:model-comparison}
\setlength{\tabcolsep}{3pt}
\resizebox{0.48\textwidth}{!}{
\begin{tabular}{lccccc}
\hline
\textbf{Model} & \textbf{Arch.} & \textbf{Params.} & \textbf{Langs.} & \textbf{\makecell{Training\\Data}} & \textbf{\makecell{Transcript-\\Free}} \\ \hline
Qwen3-TTS~\citep{qwen3tts} & AR & 1.7B & 10 & $>$5M & \gcheck \\
LEMAS-TTS~\citep{lemas} & NAR & 0.3B & 10 & 150K & \rcross \\
MOSS-TTS~\citep{moss} & AR & 8.0B & 20 & $>$1M & \gcheck \\
Fish~Audio~S2~\citep{fishs2} & AR & 4.0B & 80 & $>$10M & \rcross \\
OmniVoice~\citep{omnivoice} & NAR & 0.8B & 600+ & 581K & \rcross \\ \hline
\textbf{\stageone} (Ours) & NAR & 0.3B & 30 & 420K & \rcross \\
\textbf{\stagetwo} (Ours) & NAR & 0.4B & 30 & 420K + 40K & \gcheck \\ \hline
\end{tabular}
}
\end{table}
\paragraph{Objective Evaluation} We report Word Error Rate (WER) and Speaker Similarity (SIM-o) as objective metrics. We use Paraformer~\citep{paraformer} for Chinese and Whisper-large-v3~\citep{whisper} for other languages to transcribe the generated speech. For speaker similarity, we compute the cosine similarity between speaker embeddings of the reference and synthesized speech, extracted using a pre-trained WavLM-Large model~\citep{wavlm} fine-tuned for speaker verification.

\paragraph{Subjective Evaluation} In multilingual settings, we care about faithful voice cloning and natural, accurate speech across languages. Therefore, instead of relying on a single Mean Opinion Score (MOS), we introduce Intelligibility Mean Opinion Score (IMOS) and Similarity Mean Opinion Score (SMOS) to separately evaluate intelligibility and speaker consistency. For each language, we randomly sample 20 utterances and recruit 10 native speakers for evaluation. Each utterance is rated on a 5-point rating scale in 1-point increments. The final scores are obtained by averaging across annotators and utterances for each language. Evaluation details can be found in Appendix~\ref{apd:subjective}.

\subsection{Intra-lingual Evaluation}
\paragraph{Seed-TTS Test Set}
As shown in Table~\ref{tab:seedtts}, our X-Voice models achieve competitive WER and SIM-o performance in both Chinese and English on Seed-TTS~Test~Set~\citep{seedtts}, with a better real-time factor (RTF) than other multilingual baselines. This 
enables faster inference and 
makes the models more practical for real-time applications.
\begin{table}[ht]
\centering
\caption{Results on Seed-TTS Test Set. The boldface indicates the best result. The * denotes
reported results from the original papers. The $\downarrow$ and $\uparrow$ means lower or higher values are better. The RTF values are measured on an NVIDIA RTX 4090 GPU with a batch size of 1.}
\label{tab:seedtts}
\resizebox{0.48\textwidth}{!}{
\begin{tabular}{lccccc}
\hline
\multirow{2}{*}{\textbf{Model}} & \multirow{2}{*}{\textbf{RTF$\downarrow$}} & \multicolumn{2}{c}{\textit{\textbf{test-zh}}} & \multicolumn{2}{c}{\textit{\textbf{test-en}}} \\ \cline{3-6} 
 &  & \textbf{WER$\downarrow$} & \textbf{SIM-o$\uparrow$} & \textbf{WER$\downarrow$} & \textbf{SIM-o$\uparrow$} \\ \hline
Qwen3-TTS & 1.754 & 0.92 & 0.77 & \textbf{1.08} & 0.71 \\
LEMAS-TTS & 0.131 & 3.34 & 0.71 & 1.49 & 0.62 \\
MOSS-TTS & 0.643 & 1.46\rlap{\textsuperscript{*}} & 0.76\rlap{\textsuperscript{*}} & 1.92\rlap{\textsuperscript{*}} & 0.69\rlap{\textsuperscript{*}} \\
Fish Audio S2 & 4.801 & 1.14 & 0.73 & 1.37 & 0.65 \\
OmniVoice & 0.198 & \textbf{0.84\rlap{\textsuperscript{*}}} & \textbf{0.78\rlap{\textsuperscript{*}}} & 1.60\rlap{\textsuperscript{*}} & \textbf{0.74\rlap{\textsuperscript{*}}} \\ \hline
F5-TTS & 0.065 & 1.74\rlap{\textsuperscript{*}} & 0.75\rlap{\textsuperscript{*}} & 1.89\rlap{\textsuperscript{*}} & 0.66\rlap{\textsuperscript{*}} \\ \hline
\textbf{\stageone} & 0.073 & 1.19 & 0.75 & 1.53 & 0.65 \\
\textbf{\stagetwo} & 0.073 & 1.28 & 0.76 & 1.30 & 0.65 \\ \hline
\end{tabular}
}
\end{table}

\paragraph{LEMAS-TTS Test Set}
We compare \modelname with LEMAS-TTS on LEMAS-TTS test set, as shown in Table~\ref{tab:lemas-results}. 
\begin{table}[ht]
\centering
\caption{Results on LEMAS-TTS Test Set.}
\label{tab:lemas-results}
\small
\resizebox{\linewidth}{!}{
\begin{tabular}{lcccc}
\hline
\textbf{Model} & \textbf{WER$\downarrow$} & \textbf{SIM-o$\uparrow$} & \textbf{WER$\downarrow$} & \textbf{SIM-o$\uparrow$} \\
\cline{2-5}
& \multicolumn{2}{c}{\textbf{zh}} & \multicolumn{2}{c}{\textbf{en}} \\
\hline
LEMAS-TTS & 2.17 & 0.788 & 1.82 & 0.726 \\
\textbf{\stageone} & \textbf{1.38} & 0.816 & {1.06} & \textbf{0.745} \\
\textbf{\stagetwo} & 1.87 & \textbf{0.817} & \textbf{0.98} & 0.710 \\
\hline
& \multicolumn{2}{c}{\textbf{de}} & \multicolumn{2}{c}{\textbf{es}} \\
\hline
LEMAS-TTS & 9.94 & 0.693 & 5.60 & 0.714 \\
\textbf{\stageone} & \textbf{7.12} & 0.727 & \textbf{2.70} & {0.734} \\
\textbf{\stagetwo} & 8.24 & \textbf{0.738} & 3.27 & \textbf{0.735} \\
\hline
& \multicolumn{2}{c}{\textbf{fr}} & \multicolumn{2}{c}{\textbf{it}} \\
\hline
LEMAS-TTS & 7.27 & 0.683 & 9.50 & 0.720 \\
\textbf{\stageone} & \textbf{5.16} & {0.716} & \textbf{4.96} & \textbf{0.736} \\
\textbf{\stagetwo} & 5.67 & \textbf{0.717} & 5.46 & \textbf{0.736} \\
\hline
& \multicolumn{2}{c}{\textbf{id}} & \multicolumn{2}{c}{\textbf{pt}} \\
\hline
LEMAS-TTS & 5.47 & 0.717 & 5.30 & \textbf{0.737} \\
\textbf{\stageone} & \textbf{4.89} & 0.751 & \textbf{5.10} & 0.722 \\
\textbf{\stagetwo} & 5.37 & \textbf{0.752} & 5.71 & 0.715 \\
\hline
& \multicolumn{2}{c}{\textbf{ru}} & \multicolumn{2}{c}{\textbf{vi}} \\
\hline
LEMAS-TTS & 10.55 & 0.734 & 13.28 & 0.675 \\
\textbf{\stageone} & \textbf{6.08} & 0.756 & \textbf{10.91} & 0.702 \\
\textbf{\stagetwo} & 6.18 & \textbf{0.763} & {12.30} & \textbf{0.705} \\
\hline
\end{tabular}
}
\end{table}
\begin{table*}[ht]
\centering
\caption{Objective Results on \modelname Test Set. GT denotes ground truth audio, and Qwen3, LEMAS, MOSS, Fish, Omni denote Qwen3-TTS, LEMAS-TTS, MOSS-TTS, Fish Audio S2, OmniVoice, respectively. The boldface and underline indicate the best and the second-best result.}
\label{tab:obj-x-voice}
\setlength{\tabcolsep}{5pt}
\resizebox{\textwidth}{!}{
\begin{tabular}{lcccccccclcccccccc}
\hline
\multicolumn{1}{c}{\multirow{2}{*}{\textbf{Language}}} & \multicolumn{8}{c}{\textbf{WER$\downarrow$}} &  & \multicolumn{8}{c}{\textbf{SIM-o$\uparrow$}} \\ \cline{2-9} \cline{11-18} 
\multicolumn{1}{c}{} & GT & Qwen3 & LEMAS & MOSS & Fish & Omni & \textbf{\stageone} & \textbf{\stagetwo} &  & GT & Qwen3 & LEMAS & MOSS & Fish & Omni & \textbf{\stageone} & \textbf{\stagetwo} \\ \hline
\multicolumn{18}{l}{\textit{Asian Languages}} \\
\textbf{Chinese} & 2.41 & \textbf{2.16} & 6.07 & 2.91 & 2.57 & \underline{2.23} & 2.86 & 2.87 &  & 0.723 & \underline{0.728} & 0.655 & 0.722 & 0.686 & \textbf{0.736} & 0.698 & 0.700 \\
\textbf{Indonesian} & 3.67 & - & 4.40 & - & 6.11 & \textbf{2.15} & 2.98 & \underline{2.53} &  & 0.670 & - & 0.604 & - & 0.596 & \textbf{0.682} & 0.644 & \underline{0.651} \\
\textbf{Japanese} & 8.06 & 6.69 & - & 12.93 & \underline{6.06} & \textbf{5.98} & 8.04 & 7.93 &  & 0.713 & \underline{0.723} & - & 0.703 & 0.658 & \textbf{0.724} & 0.682 & 0.701 \\
\textbf{Korean} & 4.54 & 3.64 & - & 3.54 & \textbf{1.37} & 3.42 & 2.42 & \underline{2.40} &  & 0.759 & \underline{0.738} & - & 0.728 & 0.702 & \textbf{0.748} & 0.723 & 0.731 \\
\textbf{Thai} & 3.98 & - & - & - & 8.47 & \textbf{2.90} & \underline{5.47} & 5.64 &  & 0.694 & - & - & - & 0.644 & \textbf{0.698} & 0.654 & \underline{0.671} \\
\textbf{Vietnamese} & 3.37 & - & 4.97 & - & 29.31 & 4.38 & \underline{4.25} & \textbf{3.67} &  & 0.741 & - & 0.642 & - & \underline{0.696} & \textbf{0.727} & 0.677 & 0.686 \\ \hline
\multicolumn{18}{l}{\textit{European Languages Widely Used in TTS}} \\
\textbf{English} & 4.63 & 3.89 & 4.15 & 3.54 & 3.01 & 2.44 & \underline{2.36} & \textbf{2.29} &  & 0.730 & \underline{0.697} & 0.560 & 0.662 & 0.622 & \textbf{0.719} & 0.586 & 0.547 \\
\textbf{French} & 7.92 & \textbf{6.65} & 8.71 & 9.05 & 7.98 & \underline{6.88} & 8.71 & 8.55 &  & 0.742 & \underline{0.724} & 0.639 & 0.699 & 0.668 & \textbf{0.740} & 0.680 & 0.680 \\
\textbf{German} & 3.51 & \textbf{2.64} & 6.55 & 3.85 & 3.45 & \underline{2.80} & 3.76 & 3.91 &  & 0.765 & \underline{0.742} & 0.666 & 0.718 & 0.689 & \textbf{0.763} & 0.698 & 0.698 \\
\textbf{Italian} & 4.44 & \textbf{2.93} & 5.39 & 6.33 & 4.60 & \underline{3.13} & 3.95 & 3.89 &  & 0.760 & \underline{0.746} & 0.655 & 0.720 & 0.692 & \textbf{0.757} & 0.703 & 0.705 \\
\textbf{Portuguese} & 3.39 & \underline{2.78} & 3.55 & 6.19 & 2.71 & \textbf{2.11} & 3.41 & 3.29 &  & 0.724 & \underline{0.711} & 0.647 & 0.679 & 0.658 & \textbf{0.720} & 0.665 & 0.661 \\
\textbf{Russian} & 3.75 & 3.16 & 3.80 & 4.82 & 3.52 & 3.12 & \textbf{2.68} & \underline{2.74} &  & 0.743 & \underline{0.742} & 0.676 & 0.718 & 0.689 & \textbf{0.744} & 0.714 & 0.723 \\
\textbf{Spanish} & 3.32 & \textbf{2.11} & 4.60 & 3.88 & 2.90 & \underline{2.28} & 2.83 & 2.89 &  & 0.762 & \underline{0.749} & 0.667 & 0.726 & 0.694 & \textbf{0.759} & 0.695 & 0.693 \\ \hline
\multicolumn{18}{l}{\textit{Other European Languages}} \\
\textbf{Bulgarian} & 12.58 & - & - & - & 25.75 & \underline{9.45} & \textbf{9.27} & 9.75 &  & 0.730 & - & - & - & 0.668 & \textbf{0.721} & 0.709 & \underline{0.716} \\
\textbf{Croatian} & 11.33 & - & - & - & 11.42 & \textbf{4.56} & \underline{4.81} & 4.84 &  & 0.813 & - & - & - & 0.744 & \textbf{0.801} & 0.782 & \underline{0.790} \\
\textbf{Czech} & 8.01 & - & - & 12.02 & 12.70 & \textbf{4.58} & 4.96 & \underline{4.84} &  & 0.721 & - & - & 0.692 & 0.644 & \textbf{0.736} & 0.702 & \underline{0.706} \\
\textbf{Danish} & 13.26 & - & - & 19.67 & 25.93 & \textbf{10.49} & 12.53 & \underline{12.16} &  & 0.702 & - & - & 0.653 & 0.613 & \textbf{0.687} & 0.669 & \underline{0.676} \\
\textbf{Dutch} & 4.10 & - & - & - & 4.38 & \textbf{2.18} & 3.19 & \underline{3.13} &  & 0.725 & - & - & - & 0.650 & \textbf{0.713} & 0.667 & \underline{0.669} \\
\textbf{Estonian} & 18.15 & - & - & - & 28.12 & 13.11 & \underline{11.52} & \textbf{11.23} &  & 0.776 & - & - & - & 0.713 & \textbf{0.758} & \underline{0.733} & 0.732 \\
\textbf{Finnish} & 8.51 & - & - & - & 11.24 & 5.31 & \textbf{4.41} & \underline{4.47} &  & 0.753 & - & - & - & 0.672 & \textbf{0.754} & 0.719 & \underline{0.722} \\
\textbf{Greek} & 10.84 & - & - & 15.45 & 24.04 & \textbf{8.89} & \underline{10.52} & 10.72 &  & 0.614 & - & - & 0.645 & 0.617 & \textbf{0.704} & \underline{0.665} & \underline{0.665} \\
\textbf{Hungarian} & 7.23 & - & - & 19.26 & 11.34 & 6.85 & \underline{5.56} & \textbf{5.42} &  & 0.731 & - & - & 0.690 & 0.666 & \textbf{0.733} & 0.700 & \underline{0.701} \\
\textbf{Latvian} & 11.39 & - & - & - & 25.24 & 8.83 & \textbf{6.95} & \underline{7.13} &  & 0.715 & - & - & - & 0.647 & \textbf{0.714} & 0.692 & \underline{0.694} \\
\textbf{Lithuanian} & 12.65 & - & - & - & 50.33 & \textbf{11.73} & \underline{12.08} & 12.57 &  & 0.727 & - & - & - & 0.653 & \textbf{0.727} & 0.696 & \underline{0.702} \\
\textbf{Maltese} & 76.06 & - & - & - & 80.43 & 70.93 & \textbf{69.44} & \underline{68.11} &  & 0.705 & - & - & - & 0.611 & \textbf{0.687} & 0.653 & \underline{0.663} \\
\textbf{Polish} & 5.12 & - & - & 8.92 & 6.12 & \textbf{3.06} & \underline{3.30} & 3.71 &  & 0.726 & - & - & \underline{0.688} & 0.657 & \textbf{0.713} & 0.679 & 0.683 \\
\textbf{Romanian} & 9.85 & - & - & - & 26.44 & 8.71 & \underline{8.65} & \textbf{8.43} &  & 0.703 & - & - & - & 0.615 & \textbf{0.708} & 0.665 & \underline{0.672} \\
\textbf{Slovak} & 12.67 & - & - & - & 18.76 & \textbf{10.59} & \underline{10.52} & 11.04 &  & 0.699 & - & - & - & 0.620 & \textbf{0.700} & 0.670 & \underline{0.676} \\
\textbf{Slovenian} & 12.21 & - & - & - & 15.57 & 8.07 & \textbf{7.62} & \underline{7.73} &  & 0.683 & - & - & - & 0.606 & \textbf{0.675} & 0.645 & \underline{0.653} \\
\textbf{Swedish} & 7.29 & - & - & 9.71 & 8.06 & \textbf{5.06} & \underline{7.01} & 7.63 &  & 0.735 & - & - & \underline{0.696} & 0.666 & \textbf{0.734} & 0.687 & 0.687 \\ \hline
\end{tabular}
}
\end{table*}

Compared to LEMAS-TTS, \modelname achieves a reduction in WER scores for most of the languages, and consistently outperforms the baseline in SIM-o scores, indicating that our revised model design yields improved performance in both intelligibility and speaker identity preservation.

\paragraph{X-Voice Multilingual Test Set}

As shown in Table~\ref{tab:obj-x-voice}, \modelname behaves robustly across 30 languages, achieving WER values close to the ground truth and competitive SIM scores. In terms of WER, \modelname outperforms open-source multilingual systems including LEMAS-TTS, Fish Audio S2, and MOSS-TTS across most of the supported languages, and yields on-par results with the commercial model Qwen3-TTS. Notably, our model achieves best performance on English and Russian. Nevertheless, in terms of speaker similarity, our model still exhibits a slight performance gap compared with Qwen3-TTS, MOSS-TTS, and the concurrent model Omnivoice .

Subjective results in Table~\ref{tab:sub-x-voice} show that \modelname performs particularly well on low-resource European languages, achieving higher intelligibility (IMOS) and speaker similarity (SMOS) than other open-source baselines. For widely used languages, \modelname achieves comparable naturalness and speaker similarity to larger-scale systems, while showing slightly lower IMOS in languages such as Japanese and Korean and reduced SMOS in languages like German and Portuguese. This indicates that \modelname generally strikes a good balance between multilingual naturalness and speaker consistency, but still faces challenges in certain linguistic settings.

After fine-tuning without reference transcripts, \stagetwo achieves comparable or even improved IMOS in several languages, while SMOS tends to slightly degrade. The decrease in SMOS can be attributed to the absence of textual conditioning, making the model less capable of reproducing speaker-specific pronunciation patterns, particularly non-canonical or accented speech. This leads to more standardized pronunciations, which reduces perceived speaker consistency.

\begin{table*}[ht]
\centering
\caption{Subjective Results on \modelname Test Set. GT denotes ground truth audio, and Qwen3, LEMAS, MOSS, Fish denote Qwen3-TTS, LEMAS-TTS, MOSS-TTS, Fish Audio S2, respectively. As OmniVoice was released after we completed our subjective evaluation, it is not included in the reported results.}
\label{tab:sub-x-voice}
\setlength{\tabcolsep}{5pt}
\resizebox{\textwidth}{!}{
\begin{tabular}{lccccccclccccccc}
\hline
\multicolumn{1}{c}{\multirow{2}{*}{\textbf{Language}}} & \multicolumn{7}{c}{\textbf{IMOS$\uparrow$}} &  & \multicolumn{7}{c}{\textbf{SMOS$\uparrow$}} \\ \cline{2-8} \cline{10-16} 
\multicolumn{1}{c}{} & GT & Qwen3 & LEMAS & MOSS & Fish & \textbf{\stageone} & \textbf{\stagetwo} &  & GT & Qwen3 & LEMAS & MOSS & Fish & \textbf{\stageone} & \textbf{\stagetwo} \\ \hline
\textbf{Chinese} & 3.91 & \textbf{4.69} & 3.54 & 3.97 & \underline{4.40} & 4.33 & 4.33 &  & 3.53 & \textbf{4.15} & 2.92 & 3.87 & 3.83 & 3.97 & \underline{4.02} \\
\textbf{Indonesian} & 4.40 & - & \underline{4.52} & - & 3.50 & \textbf{4.60} & 4.10 &  & 4.04 & - & 3.92 & - & 3.34 & \underline{3.98} & \textbf{4.04} \\
\textbf{Japanese} & 4.68 & \textbf{4.46} & - & 3.00 & \underline{4.16} & 3.52 & 3.10 &  & 4.34 & \textbf{4.54} & - & 2.58 & \underline{4.16} & 2.72 & 2.76 \\
\textbf{Korean} & 4.30 & \textbf{4.58} & - & 3.82 & \underline{4.36} & 3.84 & 3.46 &  & 4.00 & \textbf{4.40} & - & 3.36 & \underline{4.08} & 3.40 & 3.24 \\
\textbf{Thai} & 4.44 & - & - & - & 3.10 & \textbf{3.96} & \underline{3.50} &  & 4.02 & - & - & - & 3.40 & \textbf{4.18} & \underline{3.96} \\
\textbf{Vietnamese} & 4.24 & - & 4.12 & - & 2.72 & \textbf{4.46} & \underline{4.36} &  & 4.22 & - & \textbf{3.90} & - & 2.64 & 3.78 & \underline{3.84} \\ \hline
\textbf{English} & 4.35 & \textbf{4.80} & 4.27 & 4.55 & \textbf{4.80} & \underline{4.63} & 4.48 &  & 4.23 & \textbf{4.60} & 3.50 & 4.23 & 4.37 & \underline{4.40} & 4.32 \\
\textbf{French} & 4.06 & \textbf{4.42} & 3.94 & 4.18 & 4.10 & \underline{4.20} & 3.94 &  & 3.88 & \textbf{4.16} & 3.82 & 3.90 & \underline{4.00} & 3.84 & 3.84 \\
\textbf{German} & 3.96 & \textbf{4.16} & 3.02 & 3.24 & 3.44 & \underline{3.60} & 3.68 &  & 3.44 & \textbf{3.90} & 2.50 & 3.18 & \underline{3.48} & 2.74 & 2.88 \\
\textbf{Italian} & 4.68 & \textbf{4.64} & 4.02 & 3.68 & 4.18 & \underline{4.26} & 4.02 &  & 4.36 & \textbf{4.62} & 3.84 & 3.68 & \underline{4.08} & 4.04 & 3.74 \\
\textbf{Portuguese} & 3.64 & 3.54 & \textbf{3.70} & 3.36 & \underline{3.64} & 3.30 & 3.62 &  & 3.54 & \textbf{3.72} & 3.20 & \underline{3.22} & 2.82 & 2.60 & 2.36 \\
\textbf{Russian} & 4.46 & \textbf{4.38} & 3.42 & 3.82 & \underline{4.16} & 3.98 & 3.70 &  & 3.76 & \textbf{4.26} & 3.08 & 3.44 & \underline{3.76} & 3.68 & 3.14 \\
\textbf{Spanish} & 4.00 & \textbf{4.32} & 3.62 & 3.92 & \underline{4.15} & 4.12 & 4.00 &  & 3.45 & \textbf{4.28} & 3.63 & 3.83 & 3.75 & 3.83 & \underline{3.98} \\ \hline
\textbf{Bulgarian} & 4.55 & - & - & - & 2.05 & \textbf{3.05} & \underline{2.78} &  & 4.45 & - & - & - & 2.08 & \textbf{3.03} & \underline{2.93} \\
\textbf{Croatian} & 4.82 & - & - & - & 1.82 & \underline{3.70} & \textbf{4.02} &  & 4.36 & - & - & - & 2.00 & \textbf{3.62} & \underline{3.54} \\
\textbf{Czech} & 4.06 & - & - & 2.76 & 1.92 & \textbf{4.12} & \underline{3.96} &  & 3.66 & - & - & 3.02 & 1.64 & \textbf{4.24} & \underline{3.92} \\
\textbf{Danish} & 4.82 & - & - & 3.08 & 2.60 & \underline{3.70} & \textbf{3.86} &  & 4.70 & - & - & 3.22 & 2.30 & \textbf{3.94} & \underline{3.50} \\
\textbf{Dutch} & 4.13 & - & - & - & 4.03 & \underline{4.43} & \textbf{4.60} &  & 4.03 & - & - & - & \textbf{3.68} & \underline{3.40} & 3.15 \\
\textbf{Estonian} & 4.52 & - & - & - & 3.62 & \underline{3.98} & \textbf{4.04} &  & 4.54 & - & - & - & 3.48 & \textbf{4.06} & \underline{3.90} \\
\textbf{Finnish} & 2.95 & - & - & - & \underline{3.40} & \textbf{3.45} & 2.98 &  & 3.08 & - & - & - & 3.05 & \underline{3.13} & \textbf{3.63} \\
\textbf{Greek} & 4.62 & - & - & \underline{3.64} & 2.20 & \textbf{3.98} & \textbf{3.98} &  & 3.78 & - & - & 3.20 & 2.52 & \underline{3.86} & \textbf{4.04} \\
\textbf{Hungarian} & 3.52 & - & - & 2.72 & 3.04 & \textbf{4.02} & \underline{3.98} &  & 3.72 & - & - & 2.84 & \underline{3.22} & \textbf{3.40} & \textbf{3.40} \\
\textbf{Latvian} & 4.56 & - & - & - & 1.44 & \underline{3.70} & \textbf{3.82} &  & 4.62 & - & - & - & 1.20 & \underline{3.76} & \textbf{3.80} \\
\textbf{Lithuanian} & 4.78 & - & - & - & 2.06 & \textbf{4.02} & \underline{3.76} &  & 4.68 & - & - & - & 1.64 & \textbf{3.48} & \underline{3.24} \\
\textbf{Maltese} & 4.38 & - & - & - & 3.68 & \underline{4.03} & \textbf{4.08} &  & 4.35 & - & - & - & 3.65 & \textbf{3.88} & \underline{3.70} \\
\textbf{Polish} & 4.22 & - & - & 3.08 & 3.20 & \textbf{4.28} & \underline{4.18} &  & 3.72 & - & - & 2.90 & 3.00 & \textbf{3.72} & \underline{3.58} \\
\textbf{Romanian} & 4.74 & - & - & - & 2.86 & \underline{4.56} & \textbf{4.64} &  & 4.20 & - & - & - & 2.46 & \underline{3.92} & \textbf{4.02} \\
\textbf{Slovak} & 3.66 & - & - & - & 2.22 & \textbf{4.20} & \underline{4.00} &  & 3.40 & - & - & - & 2.52 & \textbf{3.84} & \underline{3.74} \\
\textbf{Slovenian} & 4.18 & - & - & - & 1.73 & \textbf{3.40} & \underline{3.15} &  & 4.48 & - & - & - & 2.23 & \textbf{3.43} & \underline{3.15} \\
\textbf{Swedish} & 4.45 & - & - & 3.73 & 3.60 & \textbf{3.40} & \underline{3.38} &  & 3.55 & - & - & \underline{2.85} & \textbf{2.90} & 2.65 & 2.20 \\ \hline
\end{tabular}
}
\end{table*}

\begin{table*}[ht]
\centering
\caption{Cross-lingual WER Results on \modelname Test Set. A$\to$B denotes using language A as the prompt to synthesize speech in language B. }
\label{tab:cross-x-voice}
\setlength{\tabcolsep}{5pt}
\resizebox{\textwidth}{!}{
\begin{tabular}{ccccccccccc}
\hline
\textbf{Model} & \textbf{en$\to$it} & \textbf{it$\to$zh} & \textbf{zh$\to$ru} & \textbf{ru$\to$ko} & \textbf{ko$\to$en} & \textbf{en$\to$ko} & \textbf{ko$\to$ru} & \textbf{ru$\to$zh} & \textbf{zh$\to$it} & \textbf{it$\to$en} \\ \hline
Qwen3-TTS & \textbf{2.69} & \textbf{2.44} &  \underline{2.91} & 14.15 &  \underline{2.46} & 12.68 & 4.63 & \textbf{2.48} & \textbf{2.79} &  \underline{2.38} \\
LEMAS-TTS & 6.11 & 12.11 & 5.13 & - & - & - & - & 18.63 & 9.95 & 4.04 \\
MOSS-TTS & 7.21 & 11.64 & 5.13 &  \underline{3.64} & 9.16 & 4.06 & 10.23 & 7.52 & 5.72 & 12.4 \\
Fish Audio S2 & 9.16 & 11.04 & 4.71 & 3.85 & 3.60 & \textbf{1.71} & 4.92 & 11.17 & 9.93 & 4.04 \\
OmniVoice &  \underline{4.48} & 7.58 & 3.94 & 5.36 & 3.56 & 3.14 & 11.56 & 4.38 & 4.76 & 2.44 \\
\textbf{\stagetwo} & 4.70 &  \underline{3.11} & \textbf{2.85} & \textbf{3.00} & \textbf{2.15} & \underline{3.10} & \textbf{2.58} & \underline{3.22} & \underline{3.91} & \textbf{2.31} \\ \hline
\end{tabular}
}
\end{table*}

\subsection{Cross-lingual Evaluation}
 As shown in Table~\ref{tab:cross-x-voice}, \modelname achieves robust and consistent performance in voice cloning across different language families, obtaining the best or near-best WER in most language pairs. We also notice that source-language accents in cross-lingual synthesis can lead to ASR recognition errors, especially in Korean, resulting in higher WER. This also highlights the capability of our model to alleviate accent-induced errors and maintain high cross-lingual cloning accuracy.

\subsection{Ablation Study on Language ID Injection}\label{sec:lid}
We train \stageone using different LID injection methods, and evaluate their WER results on intra- and cross-lingual synthesis on the LEMAS test set, as shown in Table~\ref{tab:lid}.

The ablation study confirms that dual-level injection is indispensable for suppressing accent leakage, as evidenced by the WER reduction in cross-lingual tasks compared to other methods. Furthermore, FiLM-based textual modulation outperforms simple concatenation, validating its capacity to regulate phonetic articulation without overshadowing the high-entropy textual information.
\begin{table}[ht]
\small
\setlength{\tabcolsep}{2pt}
\caption{WER Results of Different LID Injection Strategies on LEMAS Test Set. Intra. denotes average WER on 10 languages supported in the test set. (F) denotes FiLM modulation, (C) denotes concatenation. }
\label{tab:lid}
\begin{tabular}{lccccc}
\hline
\multirow{2}{*}{\textbf{LID Inject Level}} & \multicolumn{5}{c}{\textbf{WER$\downarrow$}} \\ \cline{2-6} 
 & \textbf{Intra.} & \textbf{zh$\to$en} & \textbf{en$\to$zh} & \textbf{zh$\to$it} & \textbf{it$\to$zh} \\ \hline
No Injection & 5.46 & 1.11 & 6.03 & 6.65 & 7.37 \\
Text (F) & 5.28 & 1.06 & 6.05 & 5.89 & 6.85 \\
Text (F) + Time (C) & \textbf{4.94} & \textbf{0.90} & \textbf{1.87} & \textbf{3.44} & 2.93 \\
Text (C) + Time (C) & 5.49 & 0.94 & 2.01 & 3.89 & \textbf{2.88} \\ \hline
\end{tabular}
\end{table}

\subsection{Ablation Study on CFG Decoupling and Scheduling}\label{sec:cfg}
We use \stageone to evaluate the impact of different inference strategies on WER, SIM-o, and UTMOS, which are summarized in Table~\ref{tab:cfg}.

\begin{table}[ht]
\centering
\small
\caption {Average WER, SIM and UTMOS results of Different CFG Strategies on \modelname Test Set. Base denotes original CFG inference with a single CFG strength $w$. Decoupled denotes Decoupled CFG with acoustic guidance strength $w_A=2.5$ and linguistic guidance strength $w_L=4.0$.}
\label{tab:cfg}
\begin{tabular}{lccc}
\hline
\textbf{Strategy}   & \textbf{WER$\downarrow$}  & \textbf{SIM-o$\uparrow$}  & \textbf{UTMOS$\uparrow$}          \\ \hline
Base ($w=2.5$)                  & 8.85          & 0.693          & 3.207          \\
\ \  + Decay          & 8.90          & \textbf{0.694} & 3.249          \\
Base ($w=4.0$)          & 8.62          & 0.672          & 3.050          \\
\ \  + Decay          & 8.58          & 0.679          & 3.135          \\
Decoupled + Decay & 8.29          & 0.684          & 3.261          \\
\ \ + Warmup          & 8.27          & 0.684          & 3.282          \\
\ \ + A-Warmup   & \textbf{8.20} & 0.685          & \textbf{3.284} \\ \hline
\end{tabular}
\end{table}
As shown in the first four rows, standard joint CFG faces a clear trade-off between correctness and fidelity. Increasing the CFG scale from $w=2.5$ to $w=4.0$ improves the WER but leads to a significant drop in SIM-o and UTMOS. The addition of a Decay schedule partially mitigates the naturalness degradation, but the fundamental bottleneck remains. After introducing the Decoupled + A-Warmup strategy, the model achieves the lowest WER and the highest UTMOS. 

Interestingly, we observe that the highest SIM-o score is achieved by the Base ($w=2.5$) + Decay configuration. This suggests that while our decoupled strategies are effective at enforcing linguistic constraints, a conservative joint CFG scale remains superior for the most faithful preservation of the speaker’s original timbre. For applications where maximum speaker similarity is the primary objective, a lower, non-decoupled scale may be preferable, whereas our proposed phased decoupling is ideal for high-precision multilingual tasks.

\section{Conclusion}
In this paper, we presented \modelname, a highly efficient 0.4B  flow-matching foundation model for transcript-free, zero-shot multilingual voice cloning across 30 languages. By establishing a two-stage training paradigm, we demonstrate that a robust Stage-1 acoustic foundation, trained on a curated 420K-hour corpus, can serve as a high-fidelity data engine to supervise transcript-free synthesis in Stage 2. Our architectural innovations like Dual-Level Language Injection effectively mitigate the persistent challenges of cross-lingual accent leakage in flow-matching models.

Experimental results indicate that \modelname matches or exceeds the performance of industrial billion-parameter models in intelligibility and speaker similarity while maintaining a significantly faster inference speed. By open-sourcing our massive multilingual corpus, evaluation benchmarks, and full training recipes, we hope to lower the barrier for research in scalable speech generation and contribute to the democratization of high-fidelity TTS technology.

\section*{Limitations}

 Despite the robust performance of \modelname across 30 languages, several limitations remain. First, the speaker similarity in specific phonological contexts still exhibits room for improvement, suggesting that the trade-off between accent suppression and timbre preservation requires more fine-grained modeling. Second, although \modelname handles 30 languages individually, the modeling of intra-sentential code-switching remains to be further optimized.  Lastly, the reliance on high-quality synthetic data in Stage-2 highlights the ongoing need for research into purely unsupervised cross-lingual transfer.

\bibliography{custom}

\clearpage

\appendix

\section{Dataset Details}
\label{apd:dataset}
For the training set, we include Emilia~\citep{emilia} for Chinese and English, GigaSpeech~2~\citep{gigaspeech2} for Vietnamese, Thai, and Indonesian, KoreaSpeech\footnote{\url{https://huggingface.co/datasets/jp1924/KoreaSpeech}} for Korean, ReazonSpeech~\citep{reazonspeech} for Japanese, LEMAS~\citep{lemas} for Russian, Multilingual~Librispeech~\citep{mls} and Granary~\citep{granary} for Spanish, Italian, French and other European languages.  

For most European languages, utterances with a speaking rate slower than 5.0 characters per second or faster than 20.0 characters per second are excluded. For other languages, we implement the standard Interquartile Range (IQR) to exclude outliers. The lower bound is obtained by subtracting 1.5 times the interquartile range from the 25th percentile, and the upper bound by adding 1.5 times the interquartile range to the 75th percentile. We further remove duplicate utterances that appear more than 20 times. For training, we apply a DNSMOS threshold of 1.5 to filter the corpus.

For test set, most audio segments are restricted to a duration between 2.0s and 16.0s.  A minimum Root Mean Square (RMS) energy threshold of 0.02 and a minimum cosine similarity threshold of 0.6 are applied.
\section{Model Config Details}\label{apd:config}
Following the design of F5-TTS~\citep{f5tts}, both \stageone and \stagetwo employ a diffusion transformer (DiT) backbone with the following specifications: 22 layers, 16 attention heads, and 1024-dimensional text embeddings. The language embedding dimension is set to 512.

Following the design of Cross-Lingual~F5-TTS~\citep{clf5}, our Speaking Rate Predictor adopts a transformer-based architecture with 16 layers, 12 attention heads, and 512-dimensional embeddings. It is trained on a subset of the full \modelname dataset, using up to 250 hours of audio per language.

\section{Baseline Details}\label{apd:baseline}
\paragraph{Qwen3-TTS} An autoregressive model using a dual-track LM architecture for real-time synthesis. We use the Qwen3-TTS-12Hz-1.7B-Base version, and obtain the evaluation result with the official code and pretrained checkpoint\footnote{\url{https://huggingface.co/Qwen/Qwen3-TTS-12Hz-1.7B-Base/tree/main}}.
\paragraph{\textbf{LEMAS-TTS}} A flow-matching based non-autoregressive model trained on 150K hours of MMS force-aligned data. We use the GRL version which incorporates accent-adversarial training and CTC loss to mitigate cross-lingual accent issues. We obtain the evaluation result with the official code and pretrained checkpoint\footnote{\url{https://huggingface.co/LEMAS-Project/LEMAS-TTS/tree/main/pretrained_models/ckpts/multilingual_grl}}.
\paragraph{\textbf{MOSS-TTS}} A discrete-token autoregressive model built on a scalable discrete + AR + pretraining recipe. We use the standard MossTTSDelay version, which features multi‑head parallel RVQ prediction with delay‑pattern scheduling. We obtain the evaluation result with the official code and pretrained checkpoint\footnote{\url{https://huggingface.co/OpenMOSS-Team/MOSS-TTS/tree/main}}.
\paragraph{\textbf{Fish Speech S2}} An autoregressive model built on a decoder-only transformer combined with an RVQ-based audio codec, using a Dual-Autoregressive architecture. We use the standard s2 pro version and obtain the evaluation result with the official code and pretrained checkpoint\footnote{\url{https://huggingface.co/fishaudio/s2-pro/tree/main}}.
\paragraph{\textbf{OmniVoice}} A non-autoregressive model using diffusion language model-style architecture. We obtain the evaluation result with the official code and pretrained checkpoint\footnote{\url{https://huggingface.co/k2-fsa/OmniVoice/tree/main}}.
\section{Subjective Evaluation Details}\label{apd:subjective}
\paragraph{Instruction of SMOS test}

The goal of this test is to evaluate how closely the voice matches the reference speaker's timbre.

\textbf{Criteria:} Focus on voice quality, intonation style, and background consistency. Ignore differences in linguistic content or language.

\textbf{Rating Scale:}
    \begin{itemize}[itemsep=-1.5pt, topsep=5pt, partopsep=5pt]
        \item \textbf{Excellent (5):} Identical to the reference; perfect timbre match and environment consistency.
        \item \textbf{Good (4):} Very similar; timbre is clearly recognizable with minor variations in expression.
        \item \textbf{Fair (3):} Moderately similar; recognizable as the target speaker but with noticeable shifts in vocal texture.
        \item \textbf{Poor (2):} Dissimilar; the voice sounds different or the environment shift suggests a different speaker.
        \item \textbf{Bad (1):} Entirely different; mismatched gender, age, or fundamental articulatory habits.
    \end{itemize}
\paragraph{Instruction of IMOS test}

The goal of this test is to evaluate the pronunciation accuracy and prosodic naturalness of the speech.

\textbf{Criteria:} Focus on clarity, fluency, and nativeness. Ignore minor acoustic artifacts unless they impede understanding.

\textbf{Rating Scale:}
    \begin{itemize}[itemsep=-1.5pt, topsep=5pt, partopsep=5pt]
        \item \textbf{Excellent (5):} Native-like; perfect rhythm and intonation with no barriers to understanding.
        \item \textbf{Good (4):} Standard pronunciation; easy to understand with only trace amounts of unnaturalness.
        \item \textbf{Fair (3):} Intelligible but robotic; prosody feels stiff or pauses are slightly unnatural.
        \item \textbf{Poor (2):} Difficult to understand; contains mispronunciations, muddled syllables, or jarring pauses.
        \item \textbf{Bad (1):} Unintelligible; pronunciation is entirely incorrect or extremely distorted.
    \end{itemize}

\end{document}